\documentclass[%
 reprint,
superscriptaddress,
 amsmath,amssymb,
aps,
prb,
]{revtex4-2}

\usepackage{graphicx}
\usepackage{dcolumn}
\usepackage{bm}
\usepackage{hyperref}
\usepackage{enumitem}

\usepackage[caption=false]{subfig}
\usepackage{multirow}
\usepackage{graphicx}
\usepackage{dcolumn}
\usepackage{bm}
\usepackage{threeparttable}
\usepackage{color}
\usepackage[dvipsnames]{xcolor}
\usepackage{chemformula}
\usepackage{xspace}
\usepackage{orcidlink}
\usepackage{amsmath}
\usepackage{calrsfs}
\usepackage{ulem}
\usepackage{tabularx}

\begin{document}

\preprint{APS/123-QED}

\title{Electrically-gated laser-induced spin dynamics in\\magneto-electric iron garnet at room temperature}

\author{T. T. Gareev\,\orcidlink{0000-0002-1678-2444}}
 \email{gareevtimurt@gmail.com}
 \homepage{}
\affiliation{Radboud University Nijmegen, Institute for Molecules and Materials, 6525 AJ Nijmegen, The Netherlands}

\author{N. E. Khokhlov\,\orcidlink{0000-0001-7360-1163}}
\affiliation{Radboud University Nijmegen, Institute for Molecules and Materials, 6525 AJ Nijmegen, The Netherlands}

\author{L. Körber\,\orcidlink{0000-0001-8332-9669}}
\affiliation{Radboud University Nijmegen, Institute for Molecules and Materials, 6525 AJ Nijmegen, The Netherlands}

\author{A. P. Pyatakov\,\orcidlink{0000-0003-0558-9398}}
\affiliation{Faculty of Physics, Lomonosov Moscow State University, 119991 Moscow, Russia}
\affiliation{MIREA - Russian Technological University, 119454, Moscow, Russia}

\author{A. V. Kimel\,\orcidlink{0000-0002-0709-042X}}
\affiliation{Radboud University Nijmegen, Institute for Molecules and Materials, 6525 AJ Nijmegen, The Netherlands}

\date{\today}

\begin{abstract}
\textit{}
Ultrafast pump-probe imaging reveals that the efficiency of optical excitation of coherent spins waves in epitaxial iron garnet films can be effectively controlled by an external electric field at room temperature. 
Although a femtosecond laser pulse alone does not excite any pronounced coherent spin oscillations, an electrical gating with the field of 0.5 MV/m dramatically changes the outcome in a laser-induced launching of spin waves.
The effect, demonstrated under room-temperature conditions, is estimated to be orders of magnitude larger than in magnetic van der Waals semiconductors observed at 10 K.
This electrical gating of laser-induced spin dynamics enriches opto-magnonics with a new tool and thus opens up a new avenue in fundamental and applied magnonics research.

\end{abstract}

\maketitle


The ability to excite and control spin waves in magnets with the help of light opens up opportunities for  integration of spintronic or magnonic technologies with photonics and thus stimulates the development of novel opto-spintronic and opto-magnonic devices \cite{koopmans2024towards}. 
One of the main challenges in the field is a mismatch between the length-scales of a targeted bit size in spintronics or magnonics ($\sim$100 nm) and the telecom wavelength ($\sim$1 $\mu$m). For instance, using light in these technologies to control spins would require focusing it into a spot much smaller than the wavelength of light, which is quite challenging \cite{abbe1873beitrage, stelzer2000uncertainty}, though achievable in principle, e.g. in HAMR (Heat-Assisted Magnetic Recording) devices \cite{seagate2017hamr} or by using plasmonic antennas \cite{liu2015nanoscale}. 
Another possible approach to overcome this limitation is electric gating of optically induced spin excitations, which facilitates optical control of spins with spatial resolution defined by the area of the applied electric field \cite{kimel2019writing}.
Finding a material with the strongest effect of the electric field on light-spin coupling has become one of the challenges in modern magneto-electric research.
Recently, it was argued that 2D magnetic semiconductors provide an ideal platform for electric manipulation of magnetization \cite{hendriks2024electric}
It was shown that under low temperature conditions ($10$ K), applying an electric field of the order of GV/m can control an optically excited spin wave.
Here, using ultrafast imaging, we show that in an iron garnet epitaxial film, similar effects can be achieved at room temperature while applying electric fields three orders of magnitude lower.


\begin{figure}[t]
\includegraphics[width=1\columnwidth]{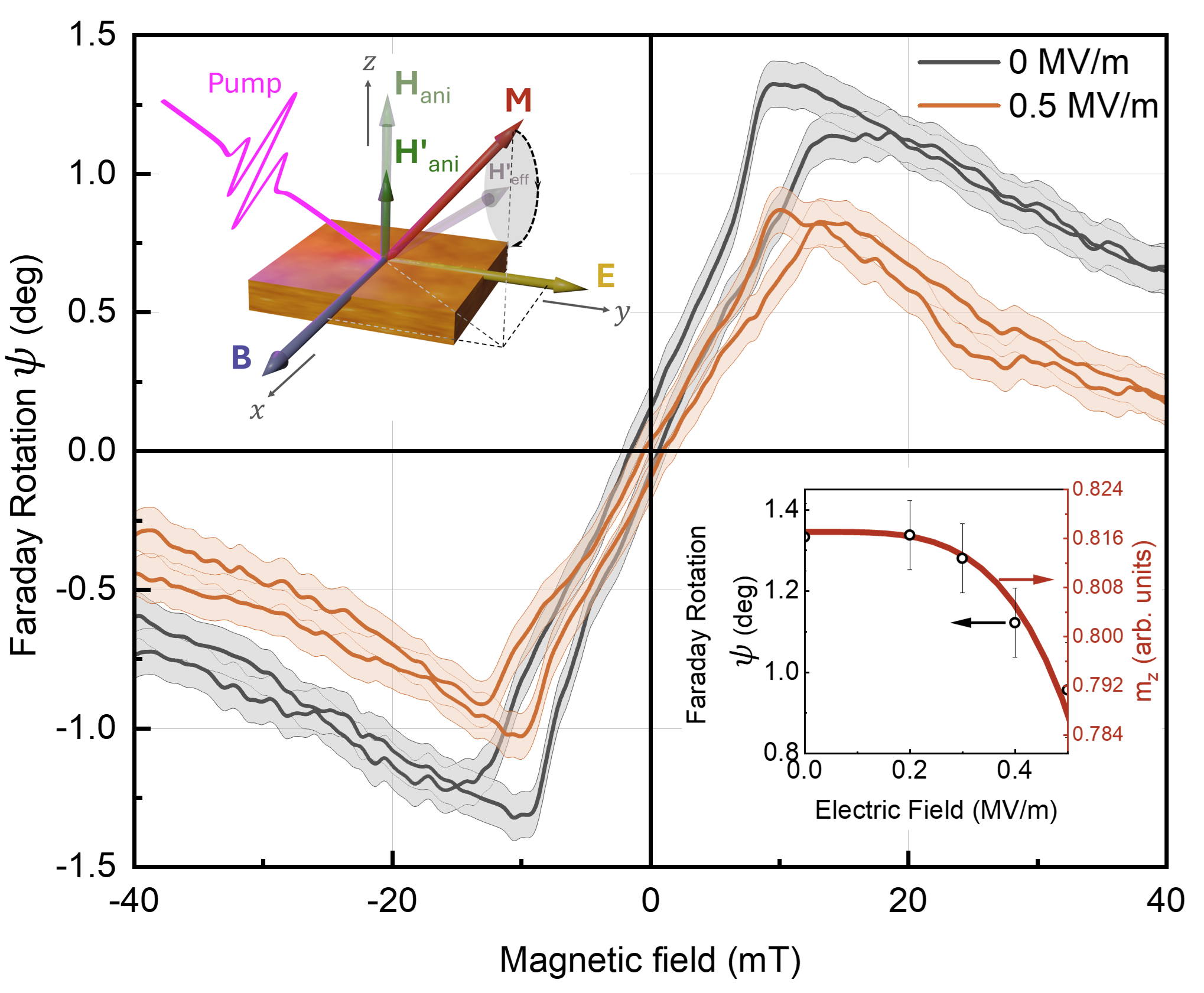}
\caption{
Magneto-optical measurements of the Faraday rotation angle $\psi$ of the probe showing the hysteresis loop without (black curve) and with the applied electric field (orange curve).
{\color{black}Shadowed areas show inaccuracy due to experimental errors.}
Top left: scheme of the experiment.
Bottom right: the maximum amplitude of the $\psi$ angle as a function of the applied electric field.
{\color{black}{Dots show the experimentally measured maximum static Faraday rotation; the solid curve shows the calculated normal component of the equilibrium magnetization $m_z$ at $t < 0$ (see details in the text)}}.
}
\label{fig:experiment_scheme}
\centering
\end{figure}

\begin{figure*}[t]
\centering
\includegraphics[width=2\columnwidth]{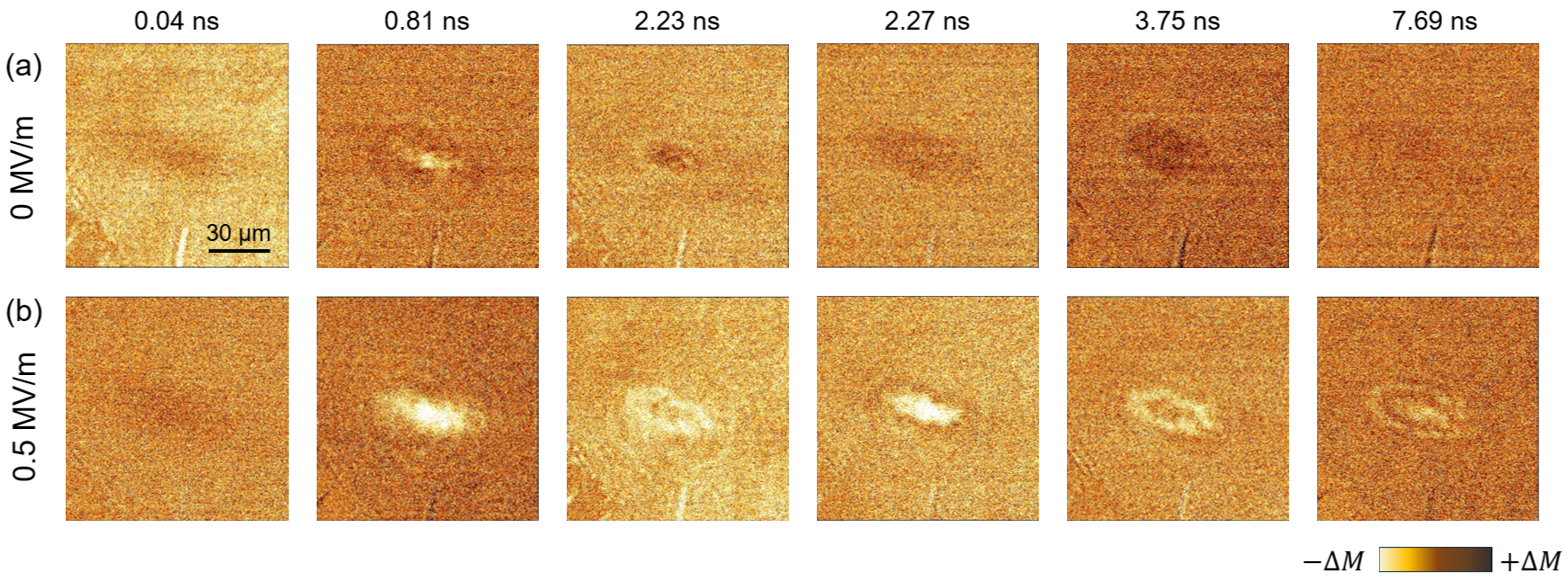}
\caption{
(a, b) Time-resolved magneto-optical images of electrically-controlled laser-induced spin dynamics.
The images were captured at different times without (a) and with (b) an applied electric field.
A magnetic field of \({B} = 30 \, \text{mT}\) was applied in all experiments. The laser fluence was \(36 \, \text{mJ/cm}^2\).
}
\label{fig:effect}
\end{figure*}

\begin{figure}[t]
\includegraphics[width=1\columnwidth]{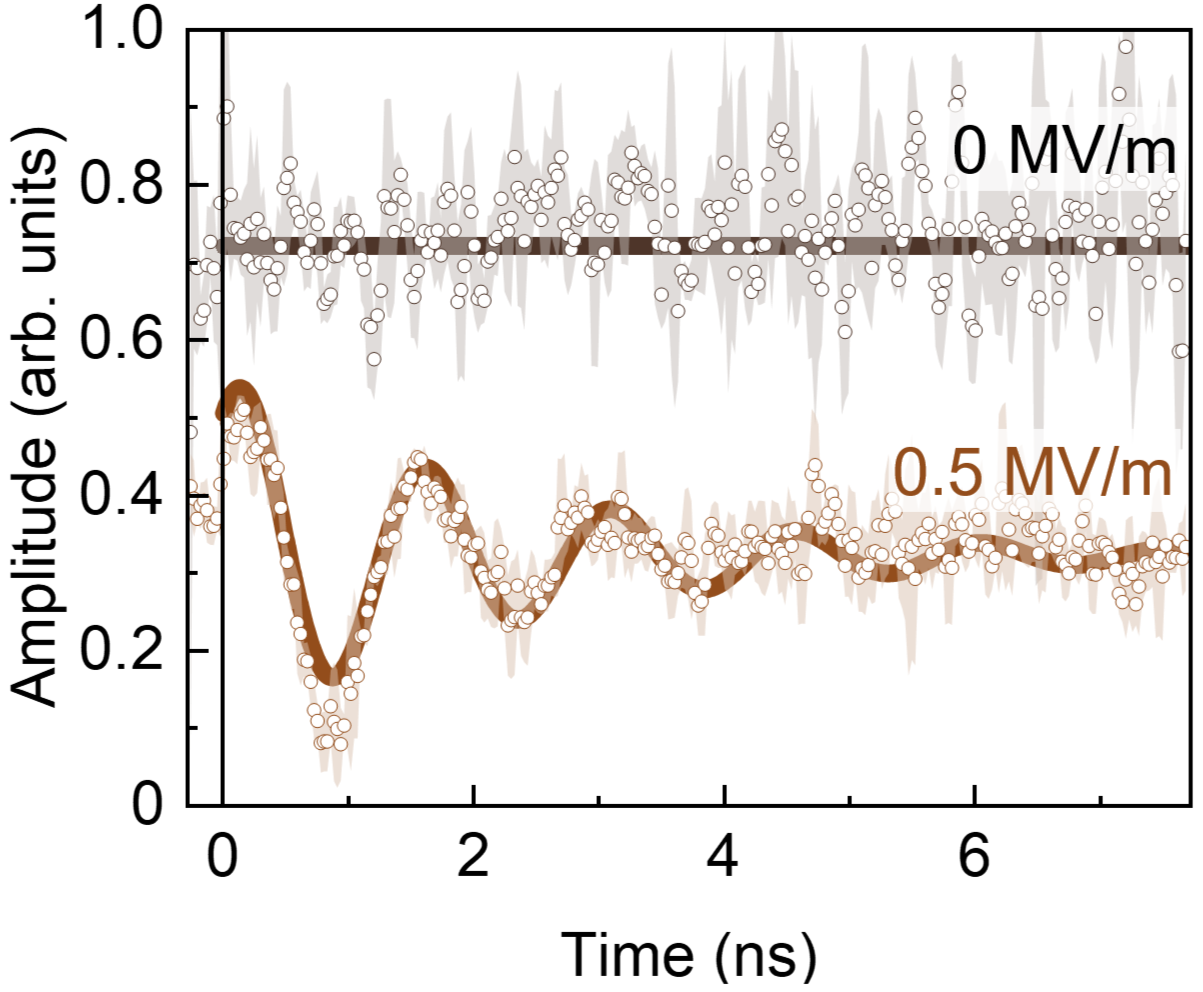}
\caption{
Magnetization dynamics extracted from the time-resolved magneto-optical images without (black) and with applied electric field $E=0.5$ MV/m (orange). 
A magnetic field $B = 30$ mT was applied in the plane of the sample. 
The laser fluence was 36 mJ/cm$^2$. 
The two sets of traces are offset relative to each other for convenience.
}
\centering
\label{fig:osc}
\end{figure}

The sample we used is an epitaxial film of iron garnet (BiLu)$_3$(FeGa)$_5$O$_{12}$ (BiLu:IG) with a thickness of 1.72 $\mu$m grown on a Gd$_3$Ga$_5$O$_{12}$ (GGG) substrate with (110) crystallographic orientation. 
In the experiments, an external magnetic field $\textbf{B}$ is applied in the sample plane with an electromagnet (top inset of Fig. \ref{fig:experiment_scheme}). 
The sample becomes in a monodomain state at $B = \pm10$ mT (Fig.\ref{fig:experiment_scheme}).
To apply the in-plane electric field $\textbf{E}$, we placed two strip electrodes: one on top of the BiLu:IG film and the second on the bottom of the GGG substrate (see Appendix \ref{appdx:exper_scheme} for details). 
The electrodes were created with silver paste and oriented so that the fields $\textbf{E}$ and $\textbf{B}$ are orthogonal.
Applying a voltage in the range of 0-500 V, we generated an electric field with a strength of up to 0.5 MV/m.
All experiments were performed at an ambient temperature of 295 K.

Magneto-optical static measurements of the Faraday rotation angle $\psi$ with $E$-field reveal changes in the hysteresis loop (bottom inset in Fig.\ref{fig:experiment_scheme}). 
Upon application of the field, the hysteresis loop shrinks.
The measurements reveal that the $E$-field works as an effective magnetic field with an in-plane orientation.
{\color{black}In particular, the $E$-field causes a tilt of the magnetization towards the sample plane, causing a shrinkage of the hysteresis loop. Such a tilt is also consistent with the calculated equilibrium positions of magnetization  shown in Fig.~\ref{fig:unit}a.}

To investigate magnetization dynamics, we performed time-resolved pump-probe imaging experiments \cite{Dolgikh2023, gareev2025strongly}. 
A Ti:Sapphire regenerative amplifier was a source of linearly polarized 45 fs laser pulses at a photon energy of 1.55\,eV with a repetition rate of 1 kHz.
The pulses are split into two parts, pump and probe. 
The most intense pulse was used as a pump and the less intense pulse was used to probe the spin dynamics.
The linearly polarized pump is focused on the sample at an incidence angle of \(10^\circ\)
in an elliptical spot that has the minor and major axes of \(30~\mu\text{m}\) and \(130~\mu\text{m}\), respectively.
The probe pulse is converted to 1.9 eV with an optical parametric amplifier. 
Then, the unfocused linearly polarized probe is used for time-resolved visualization of magnetization dynamics with a mechanical delay stage.
The Glan-Taylor polarizer is placed behind the sample in front of the CCD camera, which is used to visualize the changes in the out-of-plane magnetization component $m_z$ via the Faraday effect (for a detailed scheme of the experiment see Ref. \cite{gareev2025strongly}).


In the pump-probe experiments, we capture magneto-optical images at different moments in time, both without and with an applied $E$-field, as shown in Fig.\ref{fig:effect}a and Fig.\ref{fig:effect}b, respectively.
No pronounced spin dynamics is detected at $E=0$ (Fig.\ref{fig:effect}a). 
After the initial overlap at 0.04\,ns, a change in magneto-optical contrast is observed in the pumped area. 
At 0.81 ns, a change in phase is clearly observed. 
At 2.23 ns, the formation of a ring-like pattern is detected. However, this feature disappeared at 2.27 ns.
Finally, at 7.69 ns, no dynamics is detected in $m_z$. 
In contrast, pronounced oscillations appear when an $E$-field is applied (Fig.\ref{fig:effect}b). 
At 0.04 ns, the magneto-optical snapshot closely resembles the picture at $E=0$.
At 0.81 ns, $m_z$ changes sign, similar to the case of $E=0$, but with a significantly larger amplitude. 
At 2.23 ns, the formation of the ring-like pattern is more pronounced, with the opposite variation of $m_z$ in the central region and at the edges of the pumped area.
At 7.69 ns, multiple rings are present.
The origin of the rings can be attributed to pump-induced spatial variation of the effective anisotropy, and is discussed in more detail in our recent work \cite{gareev2025strongly}.
To gain more insight into the $E$-dependent spin dynamics, we quantify the dynamics of the average intensity in the central area of the spot and plot the value as a function of time (Fig.\,\ref{fig:osc}).
These data clearly demonstrate that the application of an electric field enables us to trigger magnetization oscillations with a frequency of 0.6 GHz. 
At the same time, no pronounced signal is observed at $E=0$. 
The full video of the spin dynamics is available in the Supplemental Materials \cite{SM}.

To explain the experimental observation of the effect of the \({E}\)-field on the  spin dynamics 
we solve the Landau-Lifshitz-Gilbert (LLG) equation \cite{landau1935theory, gilbert2004phenomenological}: 
\begin{equation}\label{eq:llg}
  \frac{d\mathbf{m}}{dt} 
  = -\frac{\gamma}{1+\alpha^2}\,\mathbf{m} \times \mathbf{B}_\text{eff}
    - \frac{\alpha\,\gamma}{1+\alpha^2}\,\mathbf{m} 
      \times \bigl(\mathbf{m} \times \mathbf{B}_\text{eff}\bigr),
\end{equation}
where $|\mathbf{m}|=1$ is the unit magnetization vector, $\gamma$ is the gyromagnetic ratio, and $\alpha$ is a dimensionless Gilbert damping parameter. 
In the experimental geometry (top inset in Fig. \ref{fig:experiment_scheme}) the effective field \textbf{$\text{B}_{\text{eff}}$} has the form 
\begin{equation}
\begin{aligned}
  \mathbf{B}_\text{eff} 
    &= B_{\text{ext}}\,\mathbf{e}_x
      +  \frac{2\xi\,|\mathbf{E}|^2}{M_s}\bigl(\mathbf{m}\cdot\mathbf{e}_y\bigr)\mathbf{m}_y  \\
    &\quad + \frac{2K_\text{eff}}{M_s} \bigl(\mathbf{m}\cdot\mathbf{e}_z\bigr)\mathbf{m}_z
\end{aligned}
\end{equation}
and includes the external magnetic field $\textit{B}_\text{ext}$, the magnetization saturation $M_s$, the magneto-electric coefficient $\xi$, and the effective anisotropy $K_\text{eff} = K_\text{uni} - \mu_{0}M_s/2$, where the uniaxial perpendicular 
magnetic anisotropy $K_\text{uni}$ is combined with the shape anisotropy, treated in the local dipole-dipole approximation. 
The \({E}\)-field is introduced as a quadratic magneto-electric term, similarly as in Ref. \cite{krichevtsov1992_MEeffect}.
We assume that the laser-induced spin dynamics arises from heating by the laser pulse, which changes the parameters $K_\text{uni}$ and $M_\text{s}$ according to the Akulov-Zener law \cite{akulov1936quantentheorie, zener1954classical, davies2019anomalously, gareev2025strongly}. 
{\color{black}We excluded non-thermal ultrafast effects (e.g. inverse Cotton–Mouton effect \cite{Kalashnikova_PRL2007}, modification of magnetic anisotropy \cite{stupakiewicz2017ultrafast}, etc.) from the model, as the detected oscillations do not demonstrate a dependence on orientation of pump polarization, and their amplitude scales with pump fluence (Suppl. Materials, Section I).}
The parameter values used for modeling are given in Appendix~\ref{appdx:sample}.

\begin{figure}[t]
    \centering
    \includegraphics[width=1
    \linewidth]{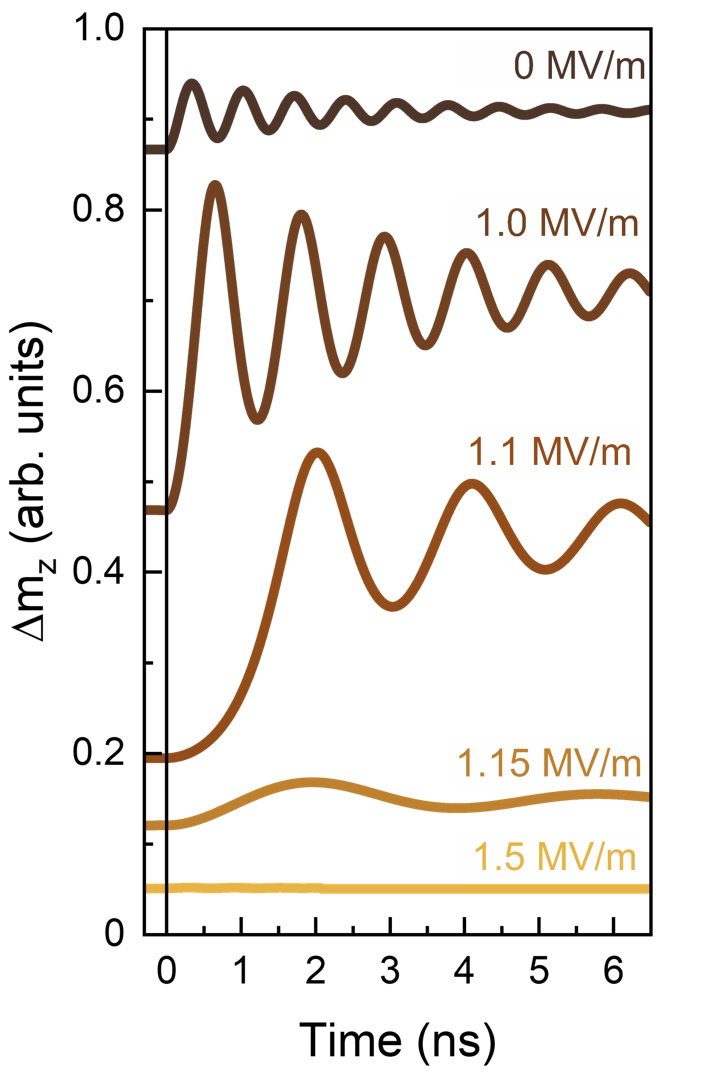}
    \caption{\
    Calculated dynamics of the out-of-plane component of the magnetization $m_z$ with and without applied electric field. 
    The dynamics is triggered as a result of ultrafast heating leading to different dynamics of magneto-crystalline anisotropy and magnetization as explained in Ref.\cite{gareev2025strongly}. 
    }
    \label{fig:model}
\end{figure}

We present numerical simulations of the time-resolved magnetization component $m_z$ under $E$-fields ranging from 0 to 1.5 MV/m (Fig.~\ref{fig:model}). {\color{black}For each value of $E$, the orientation of $\textbf{m}$ at $t<0$ is obtained by solving Eq.~\eqref{eq:llg} in a time window of 20 ns with a large damping parameter $\alpha=0.5$ and letting the system relax to its equilibrium, before perturbing the material parameters with a temperature pulse at $t = 0$.}
Although in a zero applied electric field the oscillations have a small amplitude, a rise of the $E$-field results in first an increase and then in a decrease in the amplitude with a maximum at $E = 1$ MV/m. 
From the comparison of the calculated magnetization trajectories (Fig.~\ref{fig:unit}a in Appendix~\ref{appdx:model}), it is evident that the oscillation amplitude increases, reaching a maximum with the applied field up to $E\approx1.07$ MV/m (dark orange curve in Fig.~\ref{fig:unit}b). 
Beyond this threshold, further increases in the $E$-field lead to a rapid collapse of $m_z$, causing the spin dynamics to vanish.
The simulations agree well with the experimental results within the accessible range of electric fields, limited by setup constraints.


In conclusion, we demonstrate a mechanism allowing for an effective control of the efficiency of light-spin coupling in magnets. 
Applying an electric field of the order of MV/m, we enable optical excitation of coherent spin waves in an epitaxial film of iron garnet.
Such an electrical gating facilitates a control of optically excited spin dynamics with a spatial resolution well below the diffraction limit, when coherent spin waves are excited by light only in the areas where the electric field is applied. We argue that this effect relies on a universal mechanism, which must be present in many other materials, where electric-field-induced magnetic anisotropy is strong. 
The finding practically shows that electric field is a new degree of freedom for fundamental and applied research in the fields of ultrafast magnetism, opto-spintronics and opto-magnonics. 
We demonstrate that electric‐field control in iron garnets is substantially more efficient and flexible than in 2D van der Waals magnets and similar systems: achieving comparable effects in Cr$_2$Ge$_2$Te$_6$ requires electric fields several orders of magnitude higher and cryogenic conditions \cite{hendriks2024electric}. 

In addition, apart from the dominant quadratic response, engineered asymmetry in iron garnets can produce a linear ME contribution \cite{o1967induced, krichevtsov1994linear}{\color{black}, similar to that observed in Cr$_2$O$_3$ \cite{astrov1961magnetoelectric}}, while naturally occurring inhomogeneous magneto-electric effects could be exploited
\cite{pyatakov2009flexomagnetoelectric,  khomskii2009classifying}.  
{\color{black}Nevertheless, the exact microscopic origin of the voltage-induced magnetic anisotropy hardly to be predictable for a given material \cite{dai2022review}}.
{\color{black}Given this}, artificial multiferroic heterostructures offer a powerful platform to optimize electric manipulation of the spin dynamics, where interfacial strain, spin-exchange, and charge coupling, as well as voltage-controlled magnetic anisotropy {\color{black}(e.g. in material stacks such as La$_{1-x}$Sr$_x$MnO$_3$/BaTiO$_3$, La$_{1-x}$Sr$_x$MnO$_3$/BiFeO$_3$, Fe/BaTiO$_3$ or Fe-rich CoFeB alloys)} \cite{vaz2013artificial, amiri2012voltage}, or voltage-driven inversion of magnetic anisotropy via strain-mediated coupling {\color{black}(e.g., in ferromagnetic/piezoelectric hybrids such as Ni/PZT piezoactuator structures)} \cite{weiler2009voltage}, enable reversible electric-field control of magnetism.
All of this opens new possibilities for enhancing electric control of spin dynamics in magnetic materials with magnetoelectric properties, making them promising systems for the continued study of magnetoelectric phenomena and laser-induced spin dynamics.

\section*{Acknowledgments}
The authors thank A. Stupakiewicz, L. Nowak, L. Shelukhin, and Th. Rasing for fruitful discussions, and K. Saeedi Ilkhchy and C. Berkhout for technical support. T.T.G., N.E.Kh., A.V.K. acknowledge support by the European Research Council ERC Grant Agreement No. 101054664 (SPARTACUS) and T.T.G., A.V.K. acknowledge support by the European Union’s Horizon 2020 Research and Innovation Program under Marie Skłodowska-Curie Grant Agreement No. 861300 (COMRAD). A.P.P. acknowledges the support of the Russian Science Foundation (grant No. 25-79-30019) for supporting the theoretical study of the magneto-electric effect. L.K. gratefully acknowledges funding from the Radboud Excellence Initiative.

\section*{Conflict of Interest Statement}
The authors have no conflicts to disclose.

The authors declare that this work has been published as a result of peer-to-peer scientific collaboration between researchers. The provided affiliations represent the actual addresses of the authors in agreement with their digital identifier (ORCID) and cannot be considered as a formal collaboration between the aforementioned institutions.

\section*{Author Contributions}
\textbf{Timur T. Gareev}: Investigation (experiment, equal; modeling, equal); Data Curation; Visualization (equal); Writing – original draft (equal); Writing –- review and editing (equal).
\textbf{Nikolai E. Khokhlov}: Investigation (experiment, equal); Methodology (experiment); Visualization (equal); Writing – original draft (equal); Writing –- review and editing (equal).
\textbf{Lukas K{\"o}rber}: Investigation (theory); Software, Writing –- review and editing (equal).
\textbf{Alexander P. Pyatakov}: Methodology (theoretical model); Writing –- original draft (equal).
\textbf{Alexey V. Kimel}: Conceptualization; Interpretation; Supervision; Writing –- review and editing (equal); Project Administration.

\section*{Data availability}
The raw data generated in this study and the processed data have been deposited in the Zenodo data base \cite{gareev2025datasetME}.

\bibliography{bibliography}

\section*{End Matter}
\appendix
\setcounter{figure}{0}
\renewcommand{\thefigure}{A\arabic{figure}}

\setcounter{figure}{0}
\renewcommand{\thefigure}{A\arabic{figure}}

\section{Distribution of the electric field}\label{appdx:exper_scheme}

\begin{figure}[b!]
\includegraphics[width=1\columnwidth]{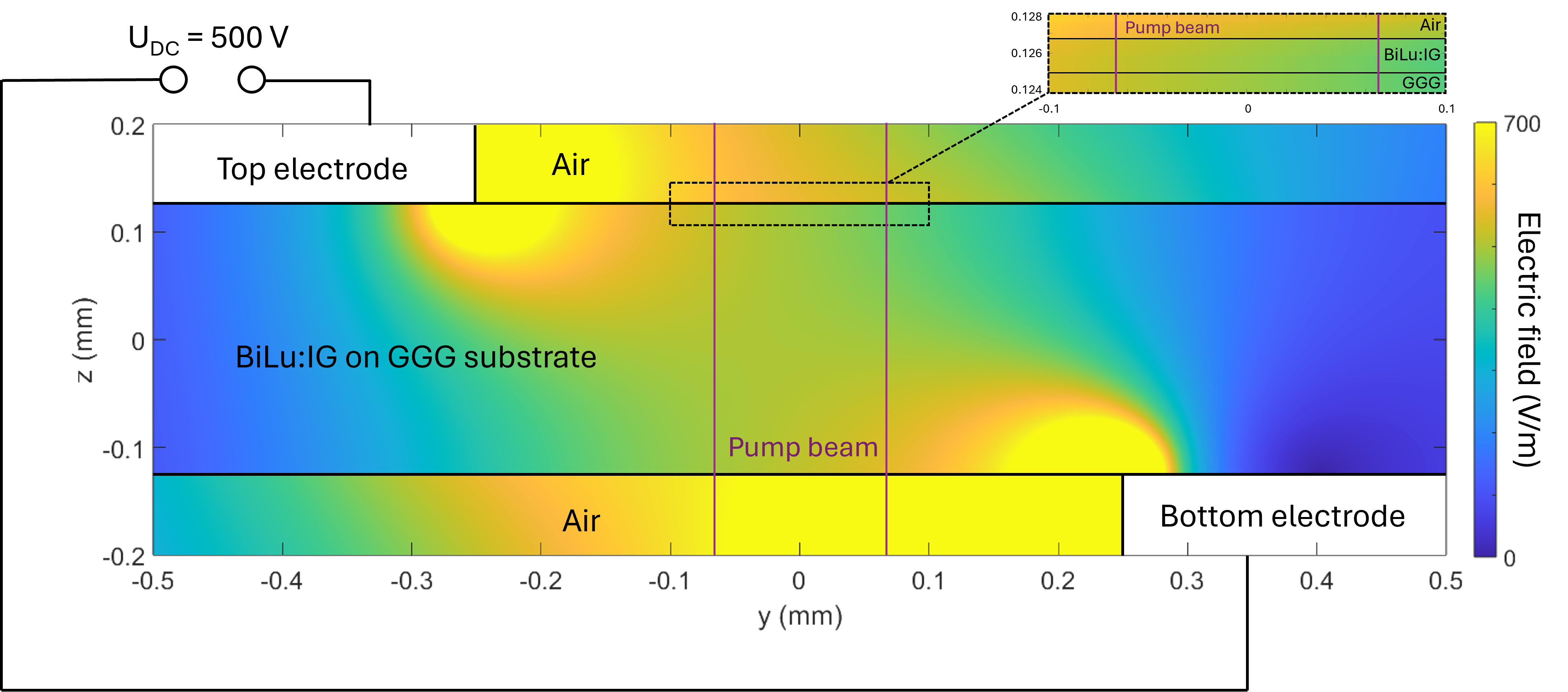}
\caption{
Calculated distribution of the electric field at applied voltage $U_{DC} = 500$ V.
The inset shows zoom in of the pumped BiLu:IG part.
Vertical purple lines indicate the pump beam.
}
\centering
\label{fig:Edistribution}
\end{figure}
The electrodes were created with silver paste on the BiLu:IG film surface and on a GGG substrate.
The distance between the electrodes along $y$-axis is 0.5 mm.
The solution of the electrostatic problem with finite element method at a voltage of 500 V gives almost in-plane orientation of the electric field inside the BiLu:IG film with a strength $E= 500$ kV/m with variation of 50 kV/m inside the pump beam (Fig.\ref{fig:Edistribution}).
The geometry provides an out-of-plane component of the electric field with a magnitude of about 4 \%.
Thus, its impact is assumed to be two orders of magnitude lower than that of the $y$-component.

\section{Magnetization dynamic}\label{appdx:model}

\begin{figure*}
\includegraphics[width=1.8\columnwidth]{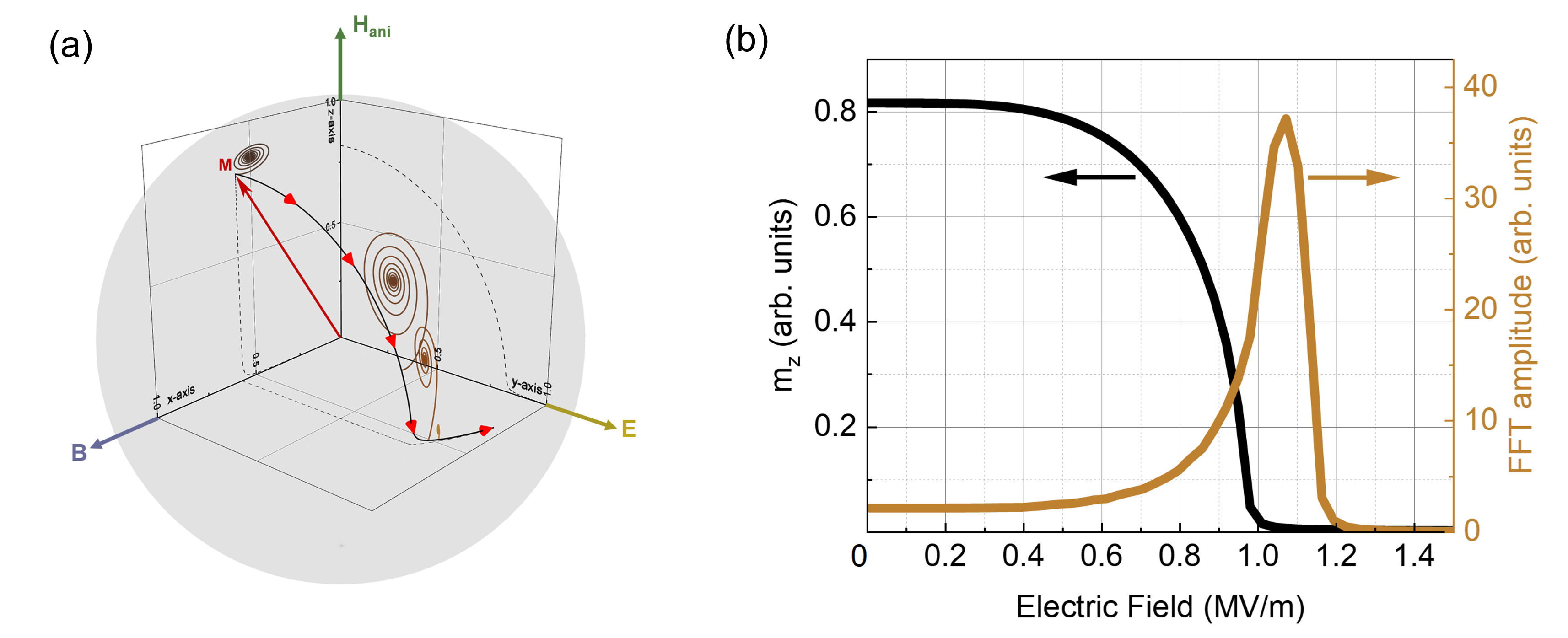}
\caption{
(a) 3D-trajectory of the magnetization vector $\textbf{m}$ at equilibrium position under applied electric field (black curve). 
The $E$-field was applied in the range $0-1.5$ MV/m.  
Trajectories of the magnetization vector $\textbf{m}$ after pump excitation under applied electric fields of 0, 1, 1.1, and 1.15 MV/m are shown as spirals on the unit sphere.
Red arrows show the direction of the changes $\textbf{m}$ under applied electric field in the equilibrium position.
(b) Calculated normal component of the equilibrium magnetization $m_z$ (black curve) and the amplitude of the laser induced spin oscillations as function of applied electric field (dark yellow curve). 
}
\centering
\label{fig:unit}
\end{figure*}

Fig.~\ref{fig:unit}a shows the trajectory of the magnetization vector $\textbf{m}$ in its equilibrium position under applied electric fields (0–1.5 kV). 
Precessional dynamics are shown for electric fields of 0, 1, 1.1, and 1.15 MV/m.

\section{Sample properties and modeling conditions}\label{appdx:sample}
The sample's magnetic parameters are $M_s = 13.5$ kA/m and $K_u = 194$ J/m$^{3}$ at 295 K.
For LLG modeling, we select parameters close to these values: $M_s(0\text{ K}) = 24$ kA/m, $K_u(0\text{ K}) = 290$ J/m$^{3}$, and $T_C = 400$ K. 
The power parameter $\beta = 12$ in the Akulov-Zener law was taken from \cite{davies2019anomalously}.
The temperature change in the model is set to $\Delta T = 20$ K.
The magnetoelectric constant was set to $\xi = 5\cdot10^{-14}$ J/(m$\cdot$V$^2$), which is of the same order as the values reported in the literature \cite{krichevtsov1992_MEeffect}.



\end{document}